# Interlayer breathing and shear modes in few-trilayer $MoS_2$ and $WSe_2$

Yanyuan Zhao,[1,†] XinLuo,[2,†] Hai Li,[3] Jun Zhang,[1] Paulo T. Araujo,[4] Chee Kwan Gan,[2] Jumiati Wu,[3] Hua Zhang,[3,*] Su Ying Quek,[2,*] Mildred S. Dresselhaus,[4,5] and Qihua Xiong[1,6,*]

[1]Division of Physics and Applied Physics, School of Physical and Mathematical Sciences, Nanyang Technological University, 21 Nanyang Link, Singapore 637371

[2]Institute of High Performance Computing, 1 Fusionopolis Way, #16-16 Connexis, Singapore 138632

[3]School of Materials Science and Engineering, Nanyang Technological University, 50 Nanyang Avenue, Singapore 639798

[4]Department of Electrical Engineering and Computer Science, Massachusetts Institute of Technology, Cambridge, Massachusetts 02139, United States of America

[5]Department of Physics, Massachusetts Institute of Technology, Cambridge, Massachusetts 02139, United States of America

[6]Division of Microelectronics, School of Electrical and Electronic Engineering, Nanyang Technological University, Singapore 639798

[†]These authors contribute to this work equally.

*To whom correspondence should be addressed. Email address: hzhang@ntu.edu.sg, queksy@ihpc.a-star.edu.sg and Qihua@ntu.edu.sg.

**Abstract**

**Two-dimensional (2D) layered transition metal dichalcogenides (TMDs) have recently attracted tremendous interest as potential valleytronic and nano-electronic materials, in addition to being well-known as excellent lubricants in the bulk. The interlayer van der Waals (vdW) coupling and low frequency phonon modes, and how they evolve with the number of layers, are important for both the mechanical and electrical properties of 2D TMDs. Here we uncover the ultra-low frequency interlayer breathing and shear modes in few-layer $MoS_2$ and $WSe_2$, prototypical layered TMDs, using both Raman spectroscopy and first principles calculations. Remarkably, the frequencies of these modes can be perfectly described using a simple linear chain model with only nearest-neighbour interactions. We show that the derived in-plane (shear) and out-of-plane (breathing) force constants from experiment remain the same from two-layer 2D crystals to the bulk materials, suggesting that the nanoscale interlayer frictional characteristics of these excellent lubricants should be independent of the number of layers.**

Bulk TMDs represent a family of about 40 layered compounds, with wide-ranging electronic properties and excellent mechanical properties as lubricants due to weak interlayer interactions.[1] Few-layer 2D TMD crystals, motivated by the experimental isolation[2] and recent scaled-up synthesis,[3-5] have been shown to have unique electronic and optical properties. For instance, the bandgap of 2D $MoS_2$ crystals exhibits an indirect-to-direct transition from a few-layer to monolayer sample;[6] while monolayer $MoS_2$ and several related TMDs have been proposed as possible valleytronics materials,[7-11] and were demonstrated as field-effect transistors.[12,13] Recent calculations predict that the carrier mobility in a $MoS_2$ monolayer is limited by optical phonon scattering due to deformation potential and Fröhlich interactions.[14] Similarly, we expect low frequency interlayer phonons to especially affect the low bias electron transport behavior via electron-phonon coupling interactions. The possible application of 2D TMDs as components in nanoscale electromechanical systems implies that a systematic understanding of their mechanical properties is required. Recently, frictional characteristics of 2D TMDs, as measured using atomic force microscopy (AFM), were found to be highly dependent on the number of layers.[15] However, the AFM measured friction between the tip and the entire 2D crystal, involving a negligible *interlayer* sliding.[15] The interlayer interactions are dominated by weak van der Waals interactions that are inherently non-local. It is thus an open and important question to understand how the interlayer interactions evolve from 3D bulk to 2D TMDs, thus elucidating the interlayer sliding contributions in friction. Therefore, probing low frequency interlayer phonon modes and interlayer force constants in 2D TMD crystals, and their evolution as a function of the number of layers, has become increasingly important.

Raman spectroscopy has been very successful in studying phonons and their couplings to electrons in 2D crystals like graphene in few-layer regime.[16] The probing of the interlayer phonons through Raman spectroscopy is challenging, since these phonon modes are usually of very low frequencies (several to tens of wavenumbers) and difficult to be distinguished from the Rayleigh background scattering. The low frequency characteristic of the interlayer phonon modes results from the weak interlayer restoring vdW force. By using a triple-grating micro Raman spectrometer and effective filters (see "Methods"), we can detect frequencies as low as ~5 $cm^{-1}$, providing a unique capability for probing low energy phonon modes. In this work, we uncover, using a combination of Raman spectroscopy and first principles calculations, the existence of *two* shear modes and *two* breathing modes in the ultra-low frequency (<55 $cm^{-1}$) region for few-layer $MoS_2$ and $WSe_2$. Similar interlayer shear modes have been identified by Raman spectra in a series of bulk layered materials such as graphite,[17] h-BN,[18] $NbSe_2$,[19] GaS,[20] $MoS_2$,[21] $WSe_2$,[22] and recently in both few-layer graphene[23] and $MoS_2$ crystals[24]. The "*organ pipe*" or "*breathing*" modes as reported here cannot be optically probed in bulk layered materials because of their optical inactivity but they might become IR or Raman active as theoretically predicted by K. H. Michel and B. Verberck in multilayer graphene and BN.[25] Although a few experimental works on few-layer graphene have reported the overtones of the interlayer breathing modes (ZO, ZA) and other high-frequency phonon modes (LO, oTO) through combination Raman scattering,[26,27] so far, the interlayer breathing modes have not yet been directly probed by any optical techniques at the low-frequency region. Here, we find that the frequencies of the experimentally observed shear modes in few-layer $MoS_2$ and $WSe_2$ redshift as the number of layers decreases, while the observed breathing modes evolve with an opposite trend. These frequencies can be perfectly described using a simple linear chain model with only nearest-neighbour interactions, with each constituent of the chain representing one layer. Using this model, we can extract both the in-plane (shear) and out-of-plane (breathing) force constants from experiment. Remarkably, these force constants remain the same from two-layer 2D crystals to the bulk materials, suggesting that the nanoscale interlayer frictional characteristics of these excellent lubricants should be independent of the number of layers.

Layered TMDs, *i.e.,* $MX_2$ (M=transition metal, X=S, Se, Te), are composed of hexagonal close-packed atomic layers. Each layer, henceforth referred to as "trilayer (TL)", consists of three atomic layers, covalently bonded to one another, and the adjacent TLs are coupled via weak vdW interactions. The $MX_2$ compounds investigated here are of the most common 2H type (atomic layers arranged in /AbA BaB/ stacking) (non-symmorphic space group $D_{6h}^{4}$ ($P6_3/mmc$)) (Figure 2a).[1] The primitive unit cell consists of two TLs (six atoms), resulting in 18 Brillouin Zone center (Γ) phonons. The irreducible representations of the phonon modes are shown as $\Gamma_{bulk} = A_{1g} + 2A_{2u} + B_{1u} + 2B_{2g} + E_{1g} + 2E_{1u} + E_{2u} + 2E_{2g}$, among which $2E_{2g}$, $E_{1g}$ and $A_{1g}$ are Raman active modes. The two-fold degenerate *E* symmetry modes represent in-plane (shear) vibrations while the *A* modes vibrate in the out-of-plane (breathing) direction along the z-axis. In few TL samples prepared by mechanical exfoliation, we believe the /AbA BaB/ stacking order is maintained. The symmetry along the *z*-axis is reduced in few-TL crystals due to the lack of translation in its direction and therefore the symmetry operations are reduced from 24 in the bulk to 12 in even- and odd-TLs each, with symmetry groups different from that for the bulk 2H materials ($D_{6h}^{4}$). The symmetry operations in few-TLs are demonstrated in Figure 1, using 1TL and 2TL as examples for odd- and even-TLs, respectively. The 12 symmetry operations in odd-TLs are: E (identity symmetry), 2 $C_3$ (the axis of the clockwise and anti-clockwise rotations is shown in Figure 1a), 3 $C_2$' (the three rotation axes are shown in Figure 1b and are lying in the $\sigma_h$ plane), $\sigma_h$ (the horizontal reflection plane is represented as the grey plane in Figure 1a), 2 $S_3$ (two $C_3$ rotations followed by a $\sigma_h$ reflection), 3 $\sigma_v$ (one of the vertical reflection planes is represented as the yellow plane in Figure 1a). Similarly, the 12 symmetry operations in even-TLs are: E, 2 $C_3$, 3 $C_2$' (the three rotation axes are shown in Figure 1e and are lying in the $\sigma_h$ plane), i (the inversion center is shown as the pink solid circle in Figure 1d), 3 $\sigma_d$ (one of the dihedral reflection planes is represented as the yellow plane in Figure 1d), 2 $S_6$ (clockwise and anti-clockwise $C_6$ rotations followed by a $\sigma_h$ reflection). Note that the $\sigma_h$ reflection alone cannot be considered as a symmetry operation in even-TLs. Consequently, odd number TL crystals belong to the symmorphic space group $D_{3h}^1$ ($P\overline{6}m2$) which has no inversion symmetry, so that the irreducible representation of the zone center phonons can be written as: $\Gamma_{odd} = \frac{3N-1}{2}\left(A'_1 + E''\right) + \frac{3N+1}{2}\left(A''_2 + E'\right)$, $N = 1, 3, 5, \ldots$ .where $N$ is the number of TLs. While even number TL crystals belong to symmorphic space group $D_{3d}^3$ ($P\overline{3}m1$) with inversion symmetry, the irreducible representation for the zone center phonons follows: $\Gamma_{even} = \frac{3N}{2}\left(A_{1g} + A_{2u} + E_g + E_u\right)$, $N = 2, 4, 6, \ldots$ . The normal mode displacements are shown in Supplementary Fig. S-2 online for 1TL, 2TL and the bulk crystal.

Let us consider the interlayer vibrations, where each TL is displaced as a whole unit. In the bulk, there are two interlayer optical phonon modes, *i.e.*, the Raman active $E_{2g}^2$ (shear mode) and the optically inactive $B_{2g}^2$ (breathing mode), where two adjacent TLs vibrate out-of-phase in-plane and out-of-plane, respectively. In *N*TL systems, there are *N*-1 two-fold degenerate interlayer shear modes and *N*-1 interlayer breathing modes. When *N* is odd, the interlayer breathing modes are either Raman-active ($A'_1$) or IR-active ($A''_2$) while the interlayer shear modes are either Raman-active ($E''$) or both Raman-active and IR-active ($E'$). When *N* is even, the interlayer breathing modes are either Raman-active ($E_g$) or IR-active ($E_u$) and the interlayer shear modes are also either Raman-active ($A_{1g}$) or IR-active ($A_{2u}$). The notations and Raman/IR activities of the phonon modes in 2H-$MoS_2$/$WSe_2$ are summarized in Table 1. The experimental observation of a given phonon mode through Raman spectroscopy depends on the symmetry selection rules as well as the scattering geometry. The Raman scattering intensity is proportional to $\left|e_i \cdot \tilde{R} \cdot e_s\right|^2$, where $e_i$ is the polarization vector of the incident light and $e_s$ is that of the scattered light. $\tilde{R}$ is the Raman tensor. A given phonon mode can be observed by Raman scattering spectroscopy only when $\left|e_i \cdot \tilde{R} \cdot e_s\right|^2$ has a nonezero value. Raman tensors of the Raman-active interlayer vibrational modes can be predicted by group theory analysis as follows:

$$A'_1 : \begin{pmatrix} a & 0 & 0 \\ 0 & a & 0 \\ 0 & 0 & b \end{pmatrix}$$
$$E' : \begin{pmatrix} c & 0 & 0 \\ 0 & -c & 0 \\ 0 & 0 & 0 \end{pmatrix}, \begin{pmatrix} 0 & d & 0 \\ d & 0 & 0 \\ 0 & 0 & 0 \end{pmatrix}$$
$$E'' : \begin{pmatrix} 0 & 0 & -c \\ 0 & 0 & 0 \\ -c & 0 & 0 \end{pmatrix}, \begin{pmatrix} 0 & 0 & 0 \\ 0 & 0 & d \\ 0 & d & 0 \end{pmatrix} \qquad (1)$$

for $MoS_2$/$WSe_2$ crystals with an odd number of TLs and

$$A_{1g} : \begin{pmatrix} a & 0 & 0 \\ 0 & a & 0 \\ 0 & 0 & b \end{pmatrix}$$
$$E_g : \begin{pmatrix} c & d & 0 \\ d & -c & 0 \\ 0 & 0 & 0 \end{pmatrix}, \begin{pmatrix} 0 & 0 & d \\ 0 & 0 & c \\ d & c & 0 \end{pmatrix} \qquad (2)$$

for even TLs.

The experimental scattering geometries of our Raman measurements can be represented by the Porto notations[28] $\bar{z}(xx)z$ and $\bar{z}(xy)z$, which correspond to parallel and perpendicular polarization configurations, respectively, in our back scattering geometry. Considering the Raman tensors of all the Raman active modes, the polarization dependence of the modes can be predicted, as summarized in the Table 1: the interlayer breathing modes $A'_1$ and $A_{1g}$ can only be observed under the $\bar{z}(xx)z$ polarization configuration (as denoted in red); the interlayer shear modes $E'$ and $E_g$ can be observed under both $\bar{z}(xx)z$ and $\bar{z}(xy)z$ polarization configurations (as denoted in blue); while the other interlayer shear mode $E''$ cannot be observed under either configuration (as denoted in black).

Few-TL $MoS_2$ and $WSe_2$ crystals are prepared by mechanical exfoliation,[2] with the thickness determined by optical contrast and atomic force microscopy measurements (see Figure 2b and Supplementary Fig. S-1 online). Figure 2c (left) shows typical anti-Stokes and Stokes Raman spectra of 1TL, 2TL, 4TL and bulk $MoS_2$ in the low frequency region (-55 to 55 $cm^{-1}$) taken with the $\bar{z}(xx)z$ polarization configuration. The spectra for the high frequency $E^1_{2g}$ and $A_{1g}$ modes (Figure 2c, right) exhibit a blueshift and redshift, respectively, from bulk to 1TL (in which these two modes actually have $E'$ and $A'_1$ vibrational symmetries, respectively), in agreement with a previous report.[29] The Raman-active bulk mode $E^2_{2g}$ (labeled as S1) corresponds to an interlayer shear mode where the adjacent TLs are vibrating out-of-phase by 180°. The S1 peak evolves to lower frequencies from bulk (~ 32 $cm^{-1}$) to 2TL (~ 22 $cm^{-1}$). Density functional theory (DFT) calculations indicate that the S1 peak corresponds in the *N*TL system to the highest frequency shear mode. We also observe a broader peak in 2TL and 4TL (labeled as B1), which can be assigned to the lowest frequency interlayer out-of-plane breathing mode by DFT. This assignment is consistent with the disappearance of B1 in the $\bar{z}(xy)z$ perpendicular polarization configuration (Figure 2d), in accordance with the Raman selection rule, where the Raman scattering intensity of a phonon mode is strictly determined by its Raman tensor and the polarization configuration of the experimental setup. Group theory predicts that under $\bar{z}(xy)z$ configuration, the Raman scattering intensity is zero for all the breathing modes and nonzero for some shear modes;

while under $\bar{z}(xx)z$ configuration, both breathing and shear modes could have nonzero Raman scattering intensities (see details in Supplementary Information-section II-3). Consequently, the out-of-plane $A_{1g}$ mode disappears in the $\bar{z}(xy)z$ configuration, while the in-plane $E_{2g}^1$ and $E_{2g}^2$ shear modes remain because of their nonzero off-diagonal matrix elements in the Raman tensors[30]. It is important to note that both the B1and S1 peaks are absent for samples with 1TL, which further confirms them as *interlayer* vibrational modes. Figures 1e-f show similar Raman spectra for $WSe_2$, with all the Raman peaks redshifted with respect to those of $MoS_2$, consistent with the larger mass per unit area in $WSe_2$. Moreover, we observed another weak peak, labeled B2, around 35 $cm^{-1}$ in 4TL $WSe_2$ under the $\bar{z}(xx)z$ configuration. DFT calculations indicate that B2 corresponds to the breathing mode with the third lowest frequency, thus explaining why it was observed only for samples thicker than 3TLs (recall that each *N*TL has (*N*-1) breathing modes).

We next conduct a systematic thickness-dependent Raman study on both materials. For the $MoS_2$ sample, as the thickness decreases from bulk to 2TL, the S1 peak redshifts from ~32 $cm^{-1}$ to ~22 $cm^{-1}$, as guided by the red dashed line (Figure 3a-b). In contrast, the second strongest peak B1 blueshifts from 9TL (~ 40 $cm^{-1}$) to 2TL (~10 $cm^{-1}$) and crosses the S1 peak at 3TL. The high frequency modes $E_{2g}^1$ and $A_{1g}$ are also dispersed as a function of TL number, scaling with a blueshift of~3 $cm^{-1}$ and a redshift of ~3.5 $cm^{-1}$, respectively, from bulk to 1TL samples (see supplementary Fig.S-9 online). The frequency dispersions of $E_{2g}^1$ and $A_{1g}$ have been proposed to determine the thickness of few-TL $MoS_2$[29], which, however, are much less sensitive than using the S1 and B1 frequencies. The B1 peak is not observed in 10TL and above, because of inadequate signal to noise ratio. Two weak peaks, labeled S2 and B2, can also be identified for samples with a thickness larger than 4 TLs, showing similar trends versus thickness with S1 and B1 (shown clearly in supplementary Fig.S-4 online–spectra before normalization). In $WSe_2$ we observe the same Raman-active modes with similar trends of evolution versus thickness (Figure 3c-d). The pronounced difference is that the B2 peak is much stronger in $WSe_2$ (Figure 3c). In both materials, the B1 and B2 modes are strongly suppressed in the $\bar{z}(xy)z$ configuration, in agreement with Raman selection rules for breathing modes, as discussed above.

The Raman spectra of these layered systems are computed from density functional perturbation theory as implemented in the Quantum-Espresso[3], within the local density approximation (LDA). Highly accurate convergence thresholds are required (see "Methods" and Supplementary Information). Our calculations indicate, as expected, that all low frequency modes (<55 $cm^{-1}$) correspond to interlayer vibrations in which each TL moves as a single unit. Furthermore, the low frequency phonon modes are essentially the same in both $MoS_2$ and $WSe_2$, except that the frequencies are lower for $WSe_2$. Figure 4a shows the two interlayer modes in the bulk material, *i.e.,* the Raman active shear mode $E_{2g}^2$ and the optically inactive breathing mode $B_{2g}^2$ . For the *N*TL systems, we obtain *N*-1 two-fold degenerate interlayer shear modes and *N*-1 interlayer breathing modes, as expected. We find that the observed S1 and S2 modes are interlayer shear modes with the *highest* and *third highest* frequencies for each *N*TL system, while the B1 and B2 modes are interlayer breathing modes with the *lowest* and *third lowest* frequencies.This explains why S1 and B1 are observed in *N*TLs with $N \geq 2$, while S2 and B2 are only observed for $N \geq 4$. We note that the breathing modes with second lowest frequencies are missing because they are Raman-inactive, belonging to the symmetry groups $A''_2$ for odd $N$ and $A_{2u}$ for even $N$ (see Table 1). Shear modes with second highest frequencies belong to the symmetry groups $E''$ for odd $N$ and $E_u$ for even $N$ and thus, cannot be observed in our Raman measurements (see Table 1).

Our DFT calculations further predict that there are, in fact, other shear and breathing modes (Fig. S6-S7 in Supplementary Information) that have the correct symmetry for observation, however, the calculated non-resonant Raman intensities, obtained within the Plazcek approximation[31,32], are essentially zero, consistent with the experimental findings. In all cases, the computed LDA frequencies match very well with experiment (Figure 5a-b), with discrepancies of ~4, ~1.5 and ~2 $cm^{-1}$ for the S1, B1 and B2 modes, respectively, for $MoS_2$ and even better agreement for $WSe_2$. Since LDA does not treat vdW interactions, we also compute the phonon frequencies using the vdW-DF functional[33,34], with Cooper's exchange interaction term.[35] Interestingly, we find that although LDA underestimates the interlayer distance in comparison with the vdW-DF functional, the vdW-DF calculation overestimates the phonon frequencies compared with experiment (see supplementary Table S-2 and Fig.S-8 online). Similar good agreement with experiment was obtained for LDA calculations on the shear mode in multilayer graphene, suggesting that although vdW-DF gives a more accurate description of the forces, LDA better describes the derivative of forces with respect to displacements.

In Figure 4b, we schematically display the normal vibrational displacements of the S1 and B1 modes in the *N*TL crystals ($2 \leq N \leq 9$). In the S1 mode, adjacent TLs are distinctly out-of-phase, while in the B1 mode, the TLs can be divided into two groups, with TLs in each group being approximately in-phase with one another. This picture is consistent with the fact that S1 is the highest frequency shear mode while B1 is the lowest frequency breathing mode, because the frequency of an interlayer phonon mode is larger if the adjacent TLs have more out-of-phase displacement. Furthermore, in the case of S1, larger $N$ implies more out-of-phase displacement between adjacent layers, leading to higher frequencies. In the case of B1, larger $N$ implies a greater proportion of approximately in-phase displacement, leading to lower frequencies. Similar arguments can be made for S2 and B2 (Figure 4c). Comparing the S1 mode with the bulk $E_{2g}^2$ mode, it is clear that the S1 mode evolves to the Raman-active $E_{2g}^2$ mode in the bulk, although the vibration amplitudes of the surface layers are slightly smaller than for layers in the middle of the few-layer materials. The reason for this difference is that the TLs at the *surface* have only one nearest neighbour TL, in contrast to TLs in the bulk that have two nearest neighbour layers. The excellent agreement between calculated and measured frequencies suggest that the TLs are weakly coupled to the silicon substrate, which is consistent with measurements obtained for suspended samples (see section V in Supplementary Information) and a recent report on frictional properties[15].

To quantify the aforementioned arguments, we consider a simple linear chain model (Figure 5c) for the interlayer modes, with each TL moving as one unit. The model further assumes that only interactions between the nearest-neighbour layers are important, and the substrate effects are neglected. The force constant $K$ is the out-of-plane constant per unit area, $K_z$, for the breathing modes, and the in-plane (shear) force constant per unit area, $K_x$, for the shear modes. A similar model has previously been used to explain the observed breathing mode in epitaxial thin films[36] and shear modes in multilayer graphene[23]. Solving this model, we obtain the eigenmodes

$$u_j^\alpha \propto \cos\left[\frac{(\alpha-1)(2j-1)\pi}{2N}\right],$$

and the corresponding phonon frequencies (in $cm^{-1}$)

$$\omega_\alpha = \sqrt{\frac{K}{2\mu\pi^2c^2}\left(1-\cos\left(\frac{(\alpha-1)\pi}{N}\right)\right)}$$

Where $j$ denotes the layer number and $\alpha$ = 1, 2, …, $N$. The $\alpha$ = 1 mode corresponds to the acoustic mode and $\alpha$ = 2, …,$N$ corresponds to the breathing modes ($K = K_z$) and shear modes ($K = K_x$), $\mu$ is the mass per unit area of the TMDs and $c$ is the speed of light in cm/s. The eigenvector $u$ is in the $z$ direction for the breathing mode and $x$ direction for the shear mode.

The aforementioned expressions fit both the measured and computed frequencies perfectly (Figure 5a-b; $\alpha = N$, $N$-2, 2, 4 for S1, S2, B1 and B2 respectively), and the resulting force constants are shown in Table2. $K_z$ derived from fits to experimental data is almost the same in both materials ($8.6\times10^{19}$ $Nm^{-3}$), while the in-plane (shear) force constant $K_x$ in $WSe_2$ is 13% larger than in $MoS_2$ ($3.1\times10^{19}$ $Nm^{-3}$ versus $2.7\times10^{19}$ $Nm^{-3}$), both are about 3 times smaller than $K_z$ and much larger than that reported in few-layer graphene ($1.28\times10^{19}$ $Nm^{-3}$).[23] We can also derive the corresponding elastic constants, $C_{33}$ and $C_{44}$, $C_{33} = K_z t$ and $C_{44} = K_x t$,[37] $t$ being the equilibrium distance between the center of each TL. This gives experimental values of $C_{33}$=52.0 GPa and $C_{44}$=16.4 GPa for $MoS_2$, and $C_{33}$=52.1 GPa and $C_{44}$=18.6 GPa for $WSe_2$.

We note that the good fits imply not only that interlayer interactions are dominated by interactions between nearest-neighbour layers, but also that the force constants $K_x$ and $K_z$ do not change significantly as $N$ is increased from 2 to 9 in $N$TLs; this is consistent with fits obtained for all computed interlayer mode frequencies with the fixed $N$ (see supplementary Fig. S-9 and S-10 online), using a single $K$ parameter. Remarkably, we note that in the limit as $N \rightarrow \infty$, the linear chain model with the above $K_x$ values predicts S1 ($E_{2g}^2$) frequencies in very good agreement with the actual calculated/measured $E_{2g}^2$ frequencies (Table 2), indicating that $K_x$ is essentially unchanged from 2TLs to the bulk material. In a recent report, AFM measurements[15] indicated that friction increased monotonically with decreasing number of layers in $MoS_2$, and this was attributed to increased elastic compliance of thinner films. We note that our results are not contradictory to, and in fact complement, these findings. As discussed by the authors,[15] there was relatively little interlayer sliding in the AFM measurements (indeed, the displacement in both in-plane and out-of-plane directions would be negligible under a ~ 10 nN lateral frictional force or normal load assuming the typical AFM tip radius of 5 nm, using the force constants we extracted). Therefore the frictional characteristics probed by the AFM experiments correspond to the limit of weak tip-layer interaction compared to interlayer interactions. However, the well-known lubricating properties of layered TMDs are in fact related to the weak *interlayer* interactions, and here, we show that these important *interlayer* interactions are essentially unchanged from 2 TLs to the bulk.

In conclusion, we have uncovered the Raman signature of both interlayer shear and breathing modes in 2H-$MoS_2$/$WSe_2$ few-TL crystals. Two breathing modes are reported for the first time, whose appearance critically depends on the polarization used in the Raman experiment. Such organ pipe breathing modes are expected to exist in many other 2D crystals. The shear and breathing modes provide effective probes of the interlayer interactions that have important implications for both mechanical and electrical properties.

## Methods

**Sample preparation.** Single- and few-TL$MoS_2$ and $WSe_2$films were isolated from bulk crystals by scotch tape-based mechanical exfoliation[2] and were then deposited onto freshly cleaned Si substrates with a 90 nm thick $SiO_2$. The layer numbers can be determined by optical contrast and thickness measurement using atomic force microscopy (Dimension 3100 with a Nanoscope IIIa controller, Veeco, CA, USA) operated in a tapping mode under ambient conditions. More details regarding sample preparation and characterization can be found in the Supplementary Information.

**Raman spectroscopy**. Raman scattering spectroscopy measurements were carried out at room temperature using a micro-Raman spectrometer (Horiba-JY T64000) equipped with a liquid nitrogen cooled charge-coupled device. The measurements were conducted in a backscattering configuration excited with a solid state green laser (λ = 532 nm). We used a reflecting Bragg grating (OptiCrate) followed by another ruled reflecting grating to filter the laser side bands, as such we can achieve ~ 5 $cm^{-1}$ limit of detection using most solid state or gas laser lines. We find our signal to noise ratio is adequate which rules out the necessity of single monochromator configuration with three notch filters as recently reported.[23] The backscattered signal was collected through a 100× objective and dispersed by a 1800 g/mm grating under a triple subtractive mode with a spectra resolution of 1 $cm^{-1}$. The laser power at the sample surface was less than 1.5 mW for $MoS_2$ and 0.3 mW for $WSe_2$. Control measurements were conducted using very low excitation power levels of 0.03 mW for both materials. No detectable difference of the peak position and full width half maximum (FWHM) intensity was observed between Raman spectra using high and low excitation power levels. Thus, the laser heating effect can be excluded in our measurements.

**Calculation details**. First principles calculations of vibrational Raman spectra are performed within density-functional theory (DFT) as implemented in the plane-wave pseudopotential code QUANTUM-ESPRESSO[3]. The local density approximation (LDA)[38] to the exchange-correlation functional is employed in the norm-conserving (NC)[39] pseudopotential throughout the calculation. For the purpose of comparison, the LDA calculations with projector-augmented wave (PAW)[40] potentials for the electron-ion interaction are performed to test the pseudopotential methods. To get converged results, plane-wave kinetic energy cutoffs of 65 Ry and 550 Ry are used for the wave functions and charge density, respectively. The slabs are separated by 16 Å of vacuum to prevent interactions between slabs (this value has been tested for convergence of phonon frequencies). A Monkhorst-Pack k-point mesh of 17×17×5 and 17×17×1 are used to sample the Brillouin Zones for the bulk and thin films systems, respectively. In the self-consistent calculation, the convergence threshold for energy is set to $10^{-9}$ eV. All the atomic coordinates and lattice constants are optimized with the Broyden–Fletcher–Goldfarb–Shanno (BFGS) quasi-Newton algorithm. During the structure optimization, the symmetry of $D_{6h}^{4}$ ($P6_3/mmc$) is imposed on the bulk, while the symmetry of $D_{3h}^{1}$ ($P\bar{6}m2$) and $D_{3d}^{3}$ ($P\bar{3}m1$) is imposed on the odd number TLs and even number TLs, respectively. The structures are considered as relaxed when the maximum component of the Hellmann-Feynman force acting on each ion is less than 0.003 eV/Å.

With the optimized structures and self-consistent wave functions, the phonon spectra and Raman intensities are calculated within density-functional perturbation theory (DFPT) as introduced by Lazzeri and Mauri.[32] For the DFPT self-consistent iteration, we used a mixing factor of 0.2 and a high convergence threshold of $10^{-18}$ eV. According to the experimental measurement, the LO-TO splitting[41,42], which results from the long range dipole-dipole interactions associated with long wavelength longitudinal phonons, are included in the calculation with the momentum vector $\mathbf{q}$ approaching zero from the $x$ direction in the dynamical matrix. We find that the LO-TO splitting does not affect the frequencies of the low frequency Raman modes reported here.

**Acknowledgements**

Q.X. gratefully acknowledges the strong support of this work from Singapore National Research Foundation through a fellowship grant (NRF-RF2009-06). This work was also supported in part by Ministry of Education via a Tier 2 grant (MOE2011-T2-2-051), start-up grant support (M58113004) and New Initiative Fund (M58110100) from Nanyang Technological University (NTU). S.Y.Q. gratefully acknowledges support from the Institute of High Performance Computing Independent Investigatorship. P.T.A. and M.S.D. acknowledge ONR-MURI-N00014-09-1-1063. H.Z. thanks the support from the Singapore National Research Foundation under a CREATE programme: Nanomaterials for Energy and Water Management, and NTU under the Start-Up Grant (M4080865.070.706022).

**Author Contributions:**

Y.Z., J.Z. and Q.X. conceived the idea. H.L., J.W. and H.Z. prepared the samples and conducted thickness measurements. Y.Z., J.Z. and Q.X. performed the Raman scattering experiments. X.L., C.K.G. and S.Y.Q. performed the first principles calculations and explained the measured and computed frequencies using the linear chain model. P.A.T.A. and M.S.D. conducted group theory analysis. All authors analyzed data and co-wrote the manuscript.

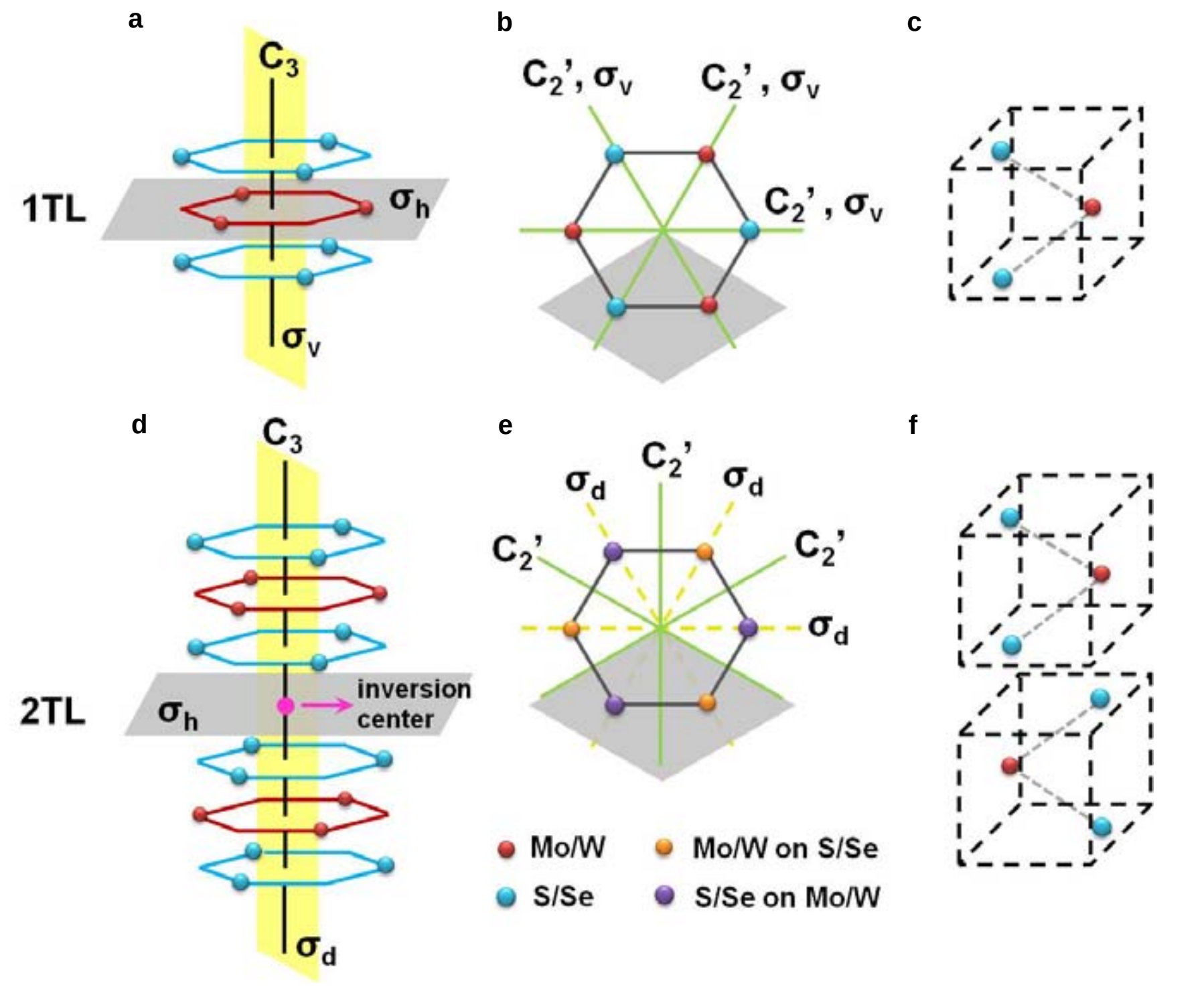

**Figure 1. Symmetry operations in 1TL and 2TL $MoS_2$/$WSe_2$. a**, side view of 1TL. The axis of the two $C_3$ operations (clockwise and anti-clockwise) is denoted as the black line. The horizontal ($\sigma_h$) and vertical ($\sigma_v$) reflection operations are shown as the grey and yellow planes, respectively. **b**, top view of 1TL. The axes of the three $C_2$' operations are denoted as the green lines, which are lying in the $\sigma_h$ plane. The top view of the $\sigma_v$ planes are also demonstrated as the green lines. The grey diamond shows the unit cell from the top view. **c**, side view of the 1TL unit cell, where one Mo atom and two S atoms are contained. **d**, side view of 2TL with /AbA BaB/ stacking. The axis of the $C_3$ operations is denoted as the black line. The inversion center is demonstrated by the pink solid circle. The horizontal ($\sigma_h$) and dihedral ($\sigma_d$) reflection operations are shown as the grey and yellow planes, respectively. Note that $\sigma_h$ is not one of the operations in the space group for 2TL. **e**, top view of 2TL. The purple spheres represent the sites where two S atoms (top TL) sit on top of one Mo atoms (bottom TL). The orange spheres represent the sites where one Mo atom (top TL) sits on top of two S atoms (bottom TL). The axes of the three $C_2$' operations are denoted as the green lines, which are lying in the $\sigma_h$ plane. The top view of the $\sigma_d$ planes are shown as the yellow dashed lines. The grey diamond shows the unit cell. **f**, side view of the 2TL unit cell, which contains four S atoms and two Mo atoms. The two TLs are represented by two dashed boxes.

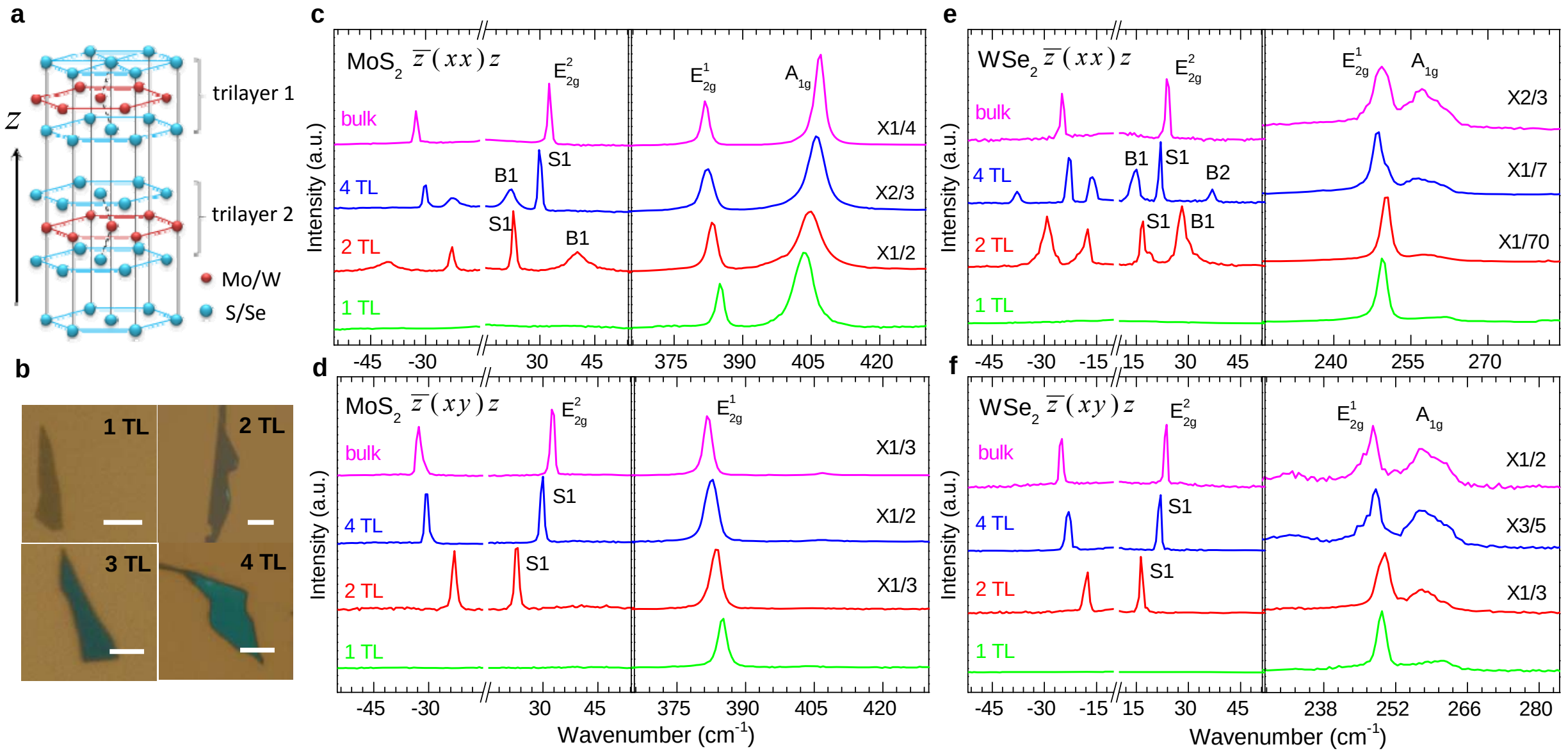

**Figure 2. Raman spectra of few-trilayer and bulk $MoS_2$ and $WSe_2$. a**,crystal structure of 2H-$MoS_2$/$WSe_2$. The primitive unit cell runs over two trilayers (TLs). **b**, Optical images of 1-4TL $MoS_2$ on 90 nm $SiO_2$/Si substrates. The scale bar is 5 μm. **c-d**, Stokes and anti-Stokes Raman spectra of 1TL, 2TL, 4TL and bulk $MoS_2$ taken under the (**c**) $\bar{z}(xx)z$ polarized back-scattering configuration and (**d**) $\bar{z}(xy)z$ polarization configuration. **e-f**, Stokes and anti-Stokes Raman spectra of 1TL, 2TL, 4TL and bulk $WSe_2$ under (**e**) $\bar{z}(xx)z$ configuration and (**f**) $\bar{z}(xy)z$ configuration. For increased clarity, all the spectra in the lowfrequency region are normalized by the intensity of the S1 peak and the spectra in the highfrequency region are normalized by the intensity of the $E_{2g}^{1}$ peak. The experimental scattering geometries of our Raman measurements are represented by the Porto notations[28] $\bar{z}(xx)z$ and $\bar{z}(xy)z$, where $\bar{z}$ and $z$ indicate wave vectors of the incident laser beam and the collected scattered light, respectively. The two alphabets in parenthesis represent the polarizations of incident and scattered light, respectively. We note that the high frequency spectra will be discussed in further detail in future manuscripts.

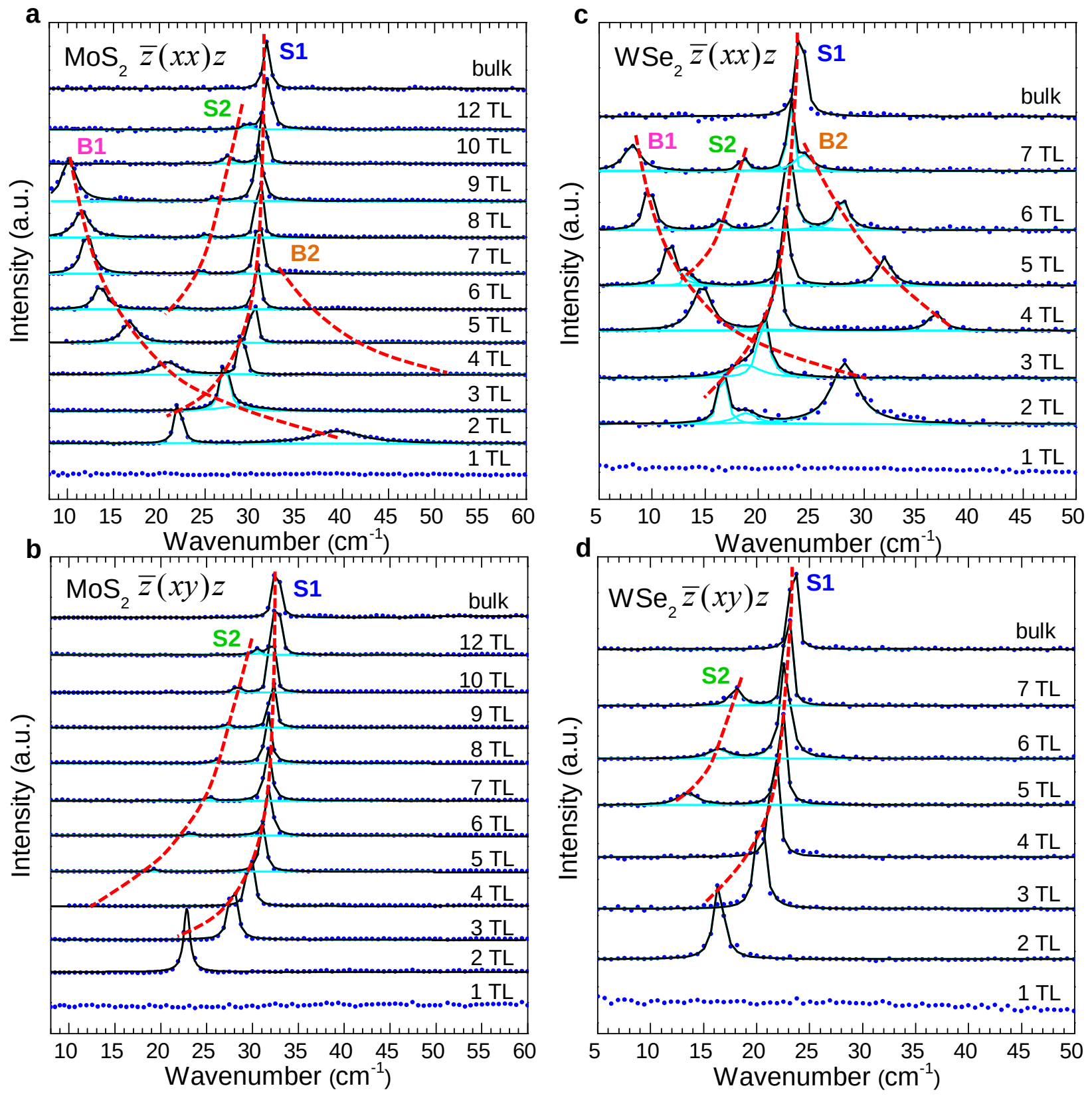

**Figure 3. Low frequency Raman spectrum evolutions as a function of trilayer number in $MoS_2$ and $WSe_2$. a-b**, Lowfrequency Raman spectra of 1-12TL $MoS_2$ measured using the (**a**) $\bar{z}(xx)z$ polarization configuration, and (**b**) the $\bar{z}(xy)z$ polarization configuration. **c-d**, Lowfrequency Raman spectra of 1-7TL $WSe_2$ measured under the (c) $\bar{z}(xx)z$ polarization configuration and (**d**) $\bar{z}(xy)z$ polarization configuration. The blue dots are experimental data points, while the black solid curves are Lorentzian fittings to the data. The Rayleigh scattering background has been subtracted for all the spectra using a polynomial baseline treatment (see Supplementary Fig. S-3 online).For increased clarity, all the spectra are normalized by the intensity of the strongest peak (corresponding to $E_{2g}^{2}$ in the bulk and labeled as S1).

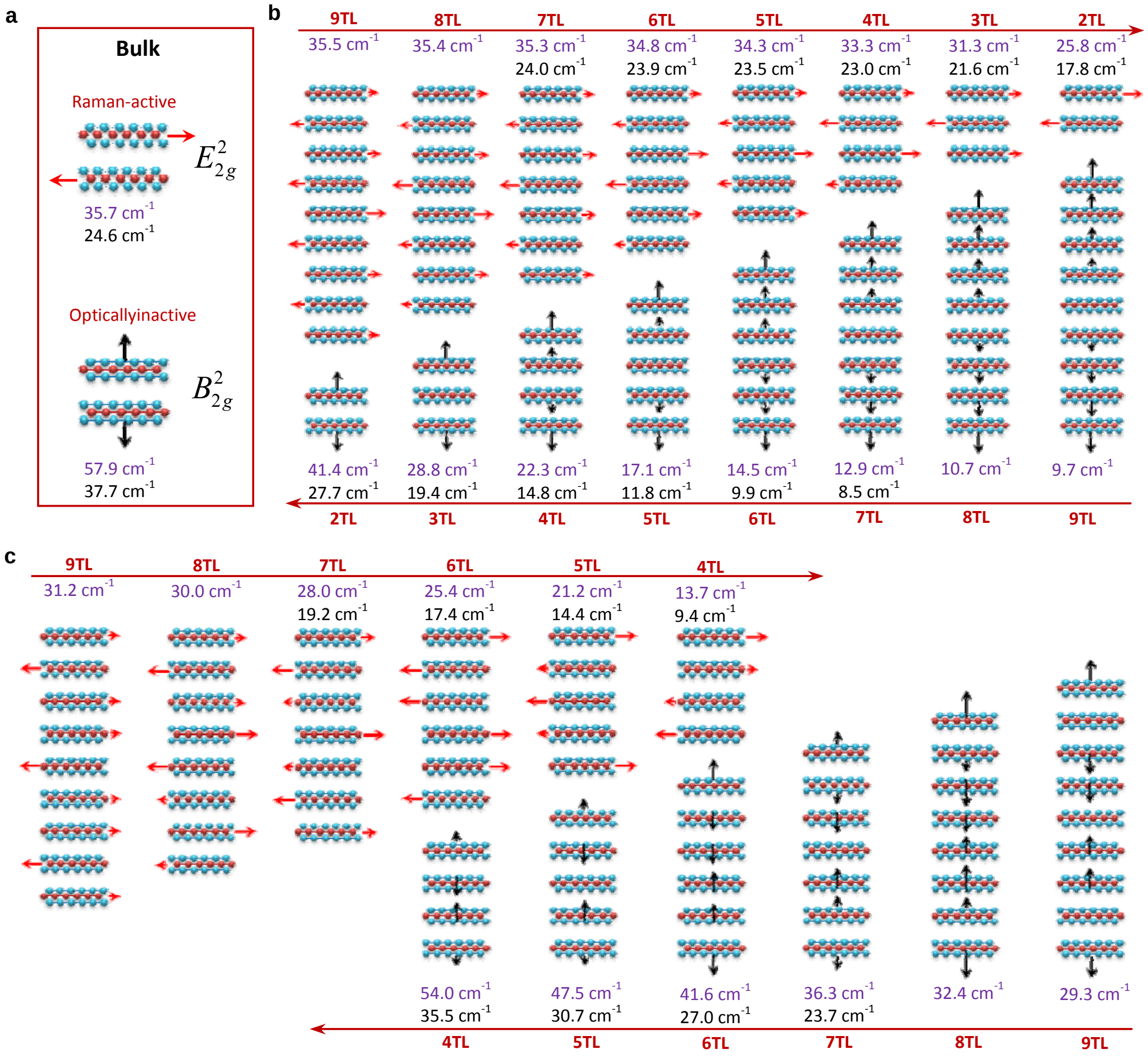

**Figure 4. Vibrational normal modes of the interlayer shear and breathing modes in $MoS_2$/$WSe_2$. a**, the vibrational normal modes of the interlayer shear ( $E_{2g}^2$ ) and breathing modes ( $B_{2g}^2$ ) in bulk 2H-$MoS_2$/$WSe_2$. The shear mode is Raman-active and the breathing mode is optically inactive. **b**, vibrational normal modes of the highestfrequency shear mode S1 (top) and the lowestfrequency breathing mode B1 (bottom) from 2TL to 9TL. **c**, vibrational normal modes of the minor shear mode S2 (top) and the minor breathing mode B2 (bottom) from 4TL to 9TL. The arrows indicate the direction of motion of the whole TL and the length of the arrows represents the magnitude. The denoted frequencies are results of the firstprinciples calculations for both $MoS_2$ (in purple) and $WSe_2$ (in black).

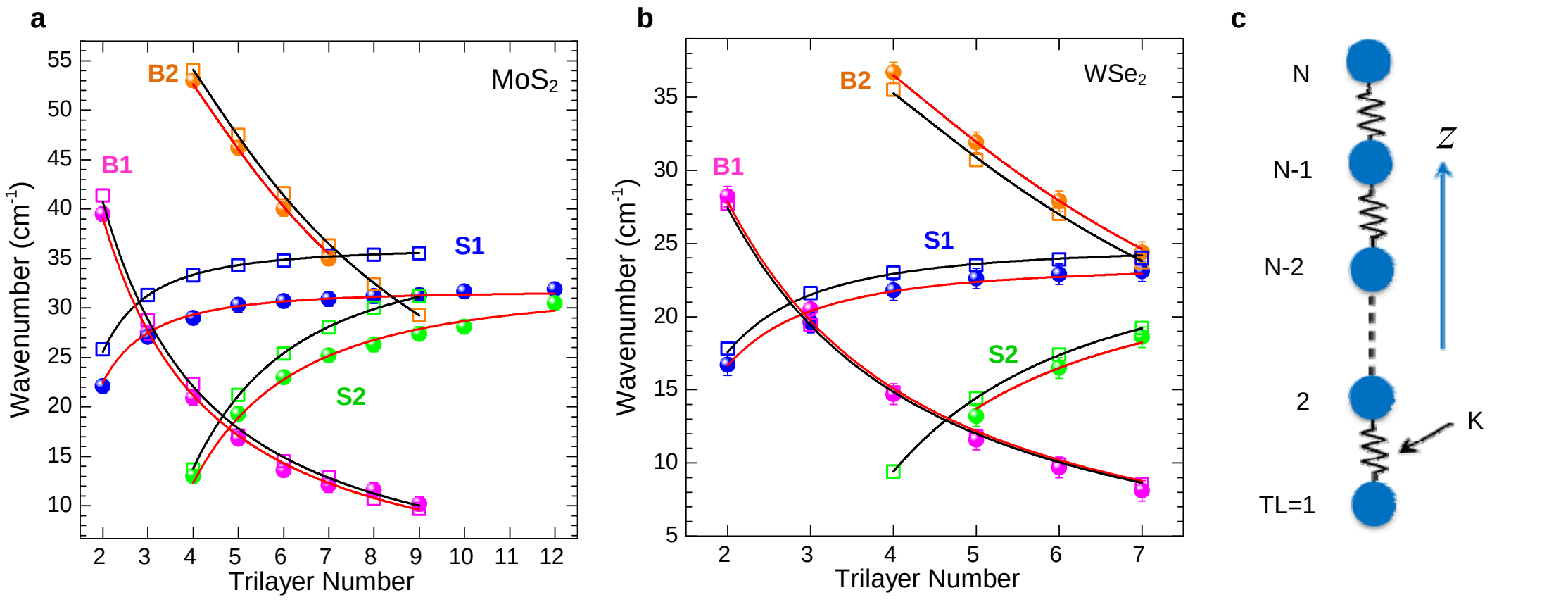

**Figure 5. Frequency evolutions of measured and computed interlayer shear and breathing modes and the linear chain model interpretation. a-b**, Plot of shear and breathing mode frequencies as a function of TL number in (**a**) $MoS_2$ and (**b**) $WSe_2$. The experimental data (solid dots) and firstprinciples calculation results (open squares) match very well. The discrepancies for S1, B1 and B2 modes in $MoS_2$ are around 4 $cm^{-1}$, 1.5 $cm^{-1}$ and 2 $cm^{-1}$, respectively. All the fitting results in **a** and **b** are shown as the red solid lines for experimental data and black solid lines for DFT results. The equations for the fits are based on the linear chain model as described in the text. **c**, schematic of the linear chain model for NTL of $MoS_2$/$WSe_2$. One blue sphere stands for a TL. The force constant is $K$ between nearest neighbour TLs.

**Table 1. Interlayer vibrational modes in bulk and few-TL $MoS_2$/$WSe_2$.** The notation and Raman/IR activity of the phonon normal modes are listed here. The phonon modes in red can be observed under both the $\bar{z}(xx)z$ and $\bar{z}(xy)z$ polarization configurations while the ones in blue can only be observed under the $\bar{z}(xx)z$ polarization configuration. The phonon modes in black cannot be observed in our Raman measurements.

| | Interlayer Shear Modes | | Interlayer Breathing Modes | |
|---|---|---|---|---|
| **Bulk** | $E_{2g}^{2}$ (R) | | $B_{2g}^{2}$ (Inactive) | |
| ***N*TL (N-odd)** | $E'$ (I+R) | $E''$ (R) | $A'_1$ (R) | $A''_2$ (I) |
| ***N*TL (N-even)** | $E_g$ (R) | $E_u$ (I) | $A_{1g}$ (R) | $A_{2u}$ (I) |

**Table 2. Force constants per unit area derived from fits to the linear chain model, and corresponding predicted S1 frequencies in the bulk.** (‡denotes the values predicted by the linear chain model, while the values in parenthesis are explicitly calculated by DFT (left column) or experimentally measured (right column) for the bulk material)

| | | DFT (LDA) | Experiment |
|---|---|---|---|
| $MoS_2$ | $K_z$($10^{19}$ $Nm^{-3}$) | 9.26 | 8.62 |
| | $K_x$($10^{19}$ $Nm^{-3}$) | 3.51 | 2.72 |
| | Bulk S1 frequency ($cm^{-1}$) | 36.1‡ (35.7) | 31.8‡ (31.7) |
| $WSe_2$ | $K_z$($10^{19}$ $Nm^{-3}$) | 8.38 | 8.63 |
| | $K_x$($10^{19}$ $Nm^{-3}$) | 3.41 | 3.07 |
| | Bulk S1 frequency ($cm^{-1}$) | 24.8‡ (24.6) | 23.5‡ (24.0) |